\documentclass[notitlepage,aps,floats,showpacs,showkeys,preprintnumbers,nofootinbib]{revtex4-1}
\pdfoutput=1

\usepackage{amsmath}
\usepackage{amssymb}
\usepackage{epsfig}
\usepackage{bbm}

\usepackage[pdftex,colorlinks=false,urlcolor=blue,linkcolor=blue]{hyperref}

\def\mueff{\mu_{\text{eff}}}
\def\fbi{~{\rm fb}^{-1}}

\def\ie{{\it i.e.}}
\def\anti{\overline}

\def\gev{~{\rm GeV}}
\def\tev{~{\rm TeV}}
\def\beq{\begin{equation}}
\def\eeq{\end{\equation}}
\def\bea{\begin{eqnarray}}
\def\eea{\end{eqnarray}}
\def\gam{\gamma}
\def\kap{\kappa}
\def\lam{\lambda}
\def\akap{A_\kap}
\def\alam{A_\lam}
\def\mhalf{m_{1/2}}

\def\mhusq{m_{H_u}^2}
\def\mhdsq{m_{H_d}^2}
\def\mssq{m_{S}^2}
\def\omghsq{\Omega h^2}
\def\amu{a_\mu}
\def\damu{\delta \amu}
\def\br{{\rm BR}}

\def\hsm{h_{\rm SM}}

\def\to{\rightarrow}
\def\hi{h_1}
\def\hii{h_2}
\def\hiii{h_3}

\def\mhi{m_{\hi}}
\def\mhii{m_{\hii}}

\def\hsm{h_{\rm SM}}

\def\tanb{\tan\beta}

\def\beq{\begin{equation}}
\def\eeq{\end{equation}}

\def\tauptaum{\tau^+\tau^-}
\def\sig{\sigma}
\def\sigres{\sig_{\text{res}}}
\def\wwww{R^h_{VBF}(WW)}
\def\ggww{R^h_{gg}(WW)}
\def\wwpp{R^h_{VBF}(\gam\gam)}
\def\ggpp{R^h_{gg}(\gam\gam)}
\def\wwbb{R^h_{VBF}(bb)}
\def\ggbb{R^h_{gg}(bb)}
\def\vev#1{\langle #1 \rangle}
\def\rts{\sqrt s}
\def\lsim{\mathrel{\raise.3ex\hbox{$<$\kern-.75em\lower1ex\hbox{$\sim$}}}}
\def\gsim{\mathrel{\raise.3ex\hbox{$>$\kern-.75em\lower1ex\hbox{$\sim$}}}}

\def\bit{\begin{itemize}}
\def\eit{\end{itemize}}
\def\bec{\begin{center}}
\def\eec{\end{center}}
\def\etc{{\it etc.}}

\begin{document}
\title{Diagnosing Degenerate Higgs Bosons at 125 GeV}

\author{John F.~Gunion}\email{jfgunion@ucdavis.edu}
\author{Yun~Jiang}\email{yunjiang@ucdavis.edu}
\affiliation{\,Department of Physics, University of California, Davis, CA 95616, USA}
\author{Sabine~Kraml}\email{sabine.kraml@lpsc.in2p3.fr}
\affiliation{\,Laboratoire de Physique Subatomique et de Cosmologie, UJF Grenoble 1, CNRS/IN2P3, INPG, 
53 Avenue des Martyrs, F-38026 Grenoble, France}


\begin{abstract}
We develop diagnostic tools that would provide incontrovertible evidence for the presence of more than one Higgs boson near 125 GeV in the LHC data.
\end{abstract}


\maketitle


Data from the ATLAS and CMS collaborations~\cite{atlashiggs,cmshiggs}
provide an essentially $5\sigma$ signal for a Higgs-like resonance with mass of order $123\mbox{--}128\gev$. 
Meanwhile, the CDF and D0 experiments have announced new results \cite{newtevatron}, based mainly on $Vh$ associated production with $h\to b\anti b$, that support the $\sim 125\gev$ Higgs-like signal.
While it is certainly possible that the observed signals in the various production/decay channels will converge towards their respective Standard Model (SM) values, the current central values for these channels deviate by about 1--2$\,\sigma$ from SM predictions. 
Clearly, a prime goal of future LHC data taking will be increased statistics, sufficient to clearly rule out or confirm a SM nature for this Higgs-like signal. Meanwhile, it is very interesting to discuss models in which the observed central values for the various channels deviate from the SM along the lines seen in the data.  
 
One of the most significant deviations in the current data is the enhancement in the $\gam\gam$ final state for both gluon fusion ($gg$) and vector boson fusion (VBF) production. Such enhancement can be obtained in a variety of models and is often associated with the observed mass eigenstate at $\sim 125\gev$ mixing with a nearby (unobserved or degenerate) state. A particularly appealing supersymmetric model that easily obtains both a Higgs mass of order $125\gev$ and significant $\gam\gam$ mode enhancements is the Next-to-Minimal Supersymmetric Standard Model (NMSSM). The NMSSM is very attractive since it solves the $\mu$ problem of the minimal supersymmetric extension of the SM (MSSM): the ad hoc parameter $\mu$ appearing in the MSSM superpotential term $\mu \hat H_u \hat H_d$ is automatically generated in the NMSSM from the $\lam \hat S \hat H_u \hat H_d$ superpotential term when the scalar component $S$ of $\hat S$ develops a VEV $\vev{S}=s$: $\mueff=\lam s$. The three CP-even Higgs fields, $H_u$, $H_d$ and $S$ mix and yield the mass eigenstates $\hi$, $\hii$ and $\hiii$.  
A $125\gev$ Higgs state with enhanced  $\gam\gam$ signal rate is easily obtained for large $\lam$ and small $\tanb$~\cite{Ellwanger:2011aa}. The $\hi$ and $\hii$ are typically close in mass in this case, with one of them being primarily the doublet-like $H_u$ while the other has a large singlet $S$ component. A particularly interesting case arises when the  $\hi$ and $\hii$ are nearly degenerate~\cite{Gunion:2012gc}. 

Given this latter possibility, a very crucial issue is  how to determine whether or not 
there are two (or more) Higgs bosons versus just one contributing to the Higgs signals at $125\gev$.  One possibility, requiring high statistics given the experimental resolution (of order $\gsim 1.5\gev$),
is that the mass peaks in the $\gam\gam$ and $4\ell$ final states would display a structure of two overlapping peaks.  However, for many of the degenerate scenarios it turns out that the $\gam\gam$ and $4\ell$ final states are dominated by only one of the degenerate Higgses, and the other one would show up primarily in $b\anti b$ and/or $\tau\tau$ final states. 
Unfortunately, mass resolutions in these channels are very poor and detection of a two peak structure using invariant mass distributions would appear to be very difficult. A direct probe of this kind of degeneracy using the full complement of final states is clearly highly desirable. 

In this Letter, we therefore develop diagnostic tools that would reveal the presence of two Higgs bosons even if they are extremely close in mass.  
We illustrate our technique 
using the NMSSM scenarios generated for \cite{Gunion:2012gc} (where NMSSM parameter ranges and all constraints are discussed in detail) in which the two lightest CP-even Higgs bosons, $\hi$ and $\hii$, both lie in the $123\mbox{--}128\gev$ mass window.  
The diagnostic tools we suggest are however fully general and can be employed for any model/scenario with degenerate Higgs-like states; to exemplify we comment via footnotes regarding differences and similarities relative to the brane model studied in \cite{Grzadkowski:2012ng} in which a Higgs and the radion mix to form two mass eigenstates, $h_1=h$, $h_2=\phi$ (or vice versa) with $m_h\sim m_\phi$.


The main production channels (denoted by $Y$) relevant for current LHC data are 
$gg\to H$ fusion ($Y=gg$) and vector boson fusion ($Y$=VBF), where VBF stands for the sum of the  $WW\to H$ and $ZZ\to H$ vector boson fusion processes. Here, $H$ stands for a generic Higgs boson. Higgs decay channels (denoted by $X$) include the high resolution $X=\gam\gam$ and $X=ZZ^*\to 4\ell$ final states, \ie\
$H\to \gam\gam$ and  $H\to ZZ^*\to 4\ell$, and the poor mass resolution $X=b\anti b$ and $X=\tauptaum$ channels. 
The most crucial production/decay channels at the LHC are $gg,{\rm VBF}\to H\to \gam\gam,4\ell$. The LHC also probes $V^*\to VH$ ($V=W$ or $Z$) with  $H \to b\anti b$,  channels for which Tevatron data is relevant, and ${\rm VBF}\to H$ with $H \to\tau^+\tau^-$.  Let us employ the notation $h_i$ for the $i^{th}$ scalar Higgs, $\hsm$ for the SM Higgs boson and $C_S^{h_i}=g_{Sh_i}/g_{S\hsm}$ is the ratio of the ${Sh_i}$ coupling  to the ${S\hsm}$ coupling, where $S=\gam\gam,gg,WW,ZZ,b b,\tauptaum$ are the cases of interest. The ratio of the 
$gg$ or VBF induced $h_i$ cross section times $\br(h_i\to X)$, relative to the corresponding value for the SM Higgs boson, takes the form
\beq
   R^{h_i}_{gg}(X)=(C_{gg}^{h_i})^2 \ {\br(h_i\to X)  \over \br(\hsm\to X)}, \quad R^{h_i}_{VBF}(X)=(C_{WW}^{h_i})^2 \ {\br(h_i\to X)  \over \br(\hsm\to X)}\,,
   \label{rdefs}
\eeq 
where the latter result assumes the custodial symmetry relation $C_{WW}^{h_i}=C_{ZZ}^{h_i}$ as applies in any doublets+singlets model; this latter also implies  $R^{h_i}_{VBF}(X)=R^{h_i}_{V^*\to V H}(X)$ and,  for either $Y=gg$ or $Y=$VBF, $R^{h_i}_Y(WW)= R^{h_i}_Y(ZZ)$. However,  if custodial symmetry is broken there are many more independent $R^{h_i}$'s.\footnote{For example, in the  Higgs-radion mixing model,  $R^{h_i}_{V^*\to V H}(X)\neq R^{h_i}_{\rm VBF}(X)$, $R^{h_i}_{gg}(WW)\neq R^{h_i}_{gg}(ZZ)$, \etc} 

In this letter we consider the case where there are two nearly degenerate Higgs bosons for which we must combine their signals.
The net signal and the effective Higgs mass, respectively,
for  given production and final decay channels $Y$ and $X$, respectively, are computed as
\beq
   R^h_Y(X)=R^{\hi}_Y(X)+ R^{\hii}_Y(X)\,,\quad
   m_h^Y(X)\equiv {R^{\hi}_{Y}(X) \mhi +R^{\hii}_Y(X) \mhii \over R^{\hi}_{Y}(X)  +R^{\hii}_Y(X) }\,.
\eeq
Of course, the extent to which it is appropriate to combine the rates from the $\hi$ and $\hii$ depends upon the degree of degeneracy and the experimental resolution, estimated to be of order $\sigres \sim 1.5\gev$~\cite{Chatrchyan:2012tw}. 
It should be noted that the widths of the $\hi$ and $\hii$ are of the same order of magnitude as the width of a 125~GeV SM Higgs boson (a few MeV), \ie\ very much smaller than this resolution.\footnote{Note that this is not an assumption. The fact that a SM-like Higgs signal in the $\gamma\gamma$ and $ZZ$ modes is even {\it visible} at the LHC tells us that the widths of any contributing Higgs boson must be very small, at most of order a few MeV as for the SM Higgs. Interference effects are negligible in this case unless one has extreme degeneracy of the two states. Our NMSSM scan points generally have $m_{h_2} -m_{h_1} > 50$ MeV for which interference effects are at most a fraction of a percent.}

As already noted, in the context of any doublets plus singlets model not all the $R^{h_i}$'s are independent; the relations among the $R^{h_i}$'s were noted above. In supersymmetric two-doublet plus singlets models we have in addition $R^{h_i}_{Y}(\tau\tau)=R^{h_i}_{Y}(bb)$. A complete independent set of $R^h$'s can be taken to be:\footnote{In other models, more (or fewer) $R^h$'s could be independent and more (or fewer) double ratios compared to those defined below could be useful/independent.   In the Higgs-radion mixing model custodial symmetry is violated, leading to {\it more} independent $R^h$'s.  For example, $R_{gg}^h(WW)\neq R_{gg}(ZZ)$ and $R_{VH}^h(X)\neq R_{VBF}^h(X)$. \label{f1}}
\beq
   \ggww,\quad \ggbb,\quad\ggpp,\quad \wwww,\quad \wwbb,\quad \wwpp\,.
\eeq
Let us now look in more detail at a given  $R^h_{Y}(X)$.  It takes the form 
\beq
   R^{h}_{Y}(X)=\sum_{i=1,2} {(C^{h_i}_Y)^2( C^{h_i}_X)^2\over C^{h_i}_\Gamma}
   \label{rsum}
\eeq
where $C^{h_i}_X$ for $X=\gam\gam,WW,ZZ,\ldots$ is the ratio of the $h_iX$ to $\hsm X$ coupling, as defined above Eq.~(\ref{rdefs}), and $C^{h_i}_{\Gamma}$ is the ratio of the total width of the $h_i$ to the SM Higgs total width.  
The diagnostic tools that we propose
to reveal the existence of a second, quasi-degenerate (but non-interfering in the small width approximation) Higgs state are the double ratios:
\beq
   \mbox{I):}~~{\wwpp/\ggpp \over \wwbb/\ggbb} ,\quad 
   \mbox{II):}~~{\wwpp/\ggpp \over \wwww/\ggww},\quad 
   \mbox{III):}~~{\wwww/\ggww \over \wwbb/\ggbb},
\label{doublerr}
\eeq
each of which should be unity if only a single Higgs boson is present  but, due to the non-factorizing nature of the sum in Eq.~(\ref{rsum}),  are generally expected to deviate from 1 if two (or more) Higgs bosons are contributing to the net $h$ signals. This occurs because the $\hi$ and $\hii$ will in general have different relative production rates in the VBF and gg fusion channels for one or more final states.  One can check that in a doublets+singlets model all other double ratios that are equal to unity for single Higgs exchange are not independent of the above three. Of course, the above three double ratios are not all independent. Which will be most useful depends upon the precision with which the $R^h$'s for different initial/final states can be measured.  For example, measurements of $R^h$ for the $bb$ final state may continue to be somewhat imprecise and it is then double ratio II) that might prove most discriminating.  Or, it could be that one of the double ratios deviates from unity by a much larger amount than the others, in which case it might be most discriminating even if the $R^h$'s involved are not measured with great precision.  

To explore how powerful these double ratios are in practice, we turn to the NMSSM scenarios with semi-unified GUT scale soft-SUSY-breaking sampled in \cite{Gunion:2012gc}.\footnote{By ``semi-unified'' we mean universal gaugino mass parameter $\mhalf$, scalar (sfermion) mass parameter $m_0$, and trilinear coupling $A_0\equiv A_t=A_b=A_\tau$ at the GUT scale, but $\mhusq$, $\mhdsq$ and $\mssq$ as well as $\alam$ and $\akap$ are taken as non-universal at $M_{\rm GUT}$.}  
These scenarios obey all experimental constraints (including $\omghsq<0.136$ and 2011 XENON100 limits on the spin-independent scattering cross section) except that 
the SUSY contribution to the anomalous magnetic moment of the muon, 
$\damu$, is too small to explain the discrepancy between the observed value $a_\mu$ and that predicted by the SM. For a full discussion of the kind of NMSSM model employed see also~\cite{Gunion:2012zd,Ellwanger:2012ke}.  

\begin{figure}[h]\centering
\includegraphics[width=0.55\textwidth]{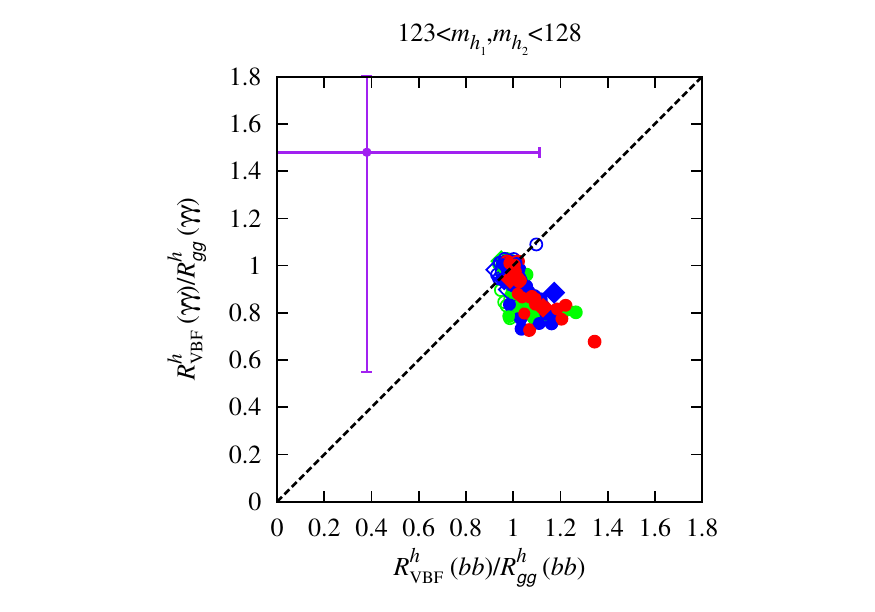}
\hspace{-30mm}
\includegraphics[width=0.55\textwidth]{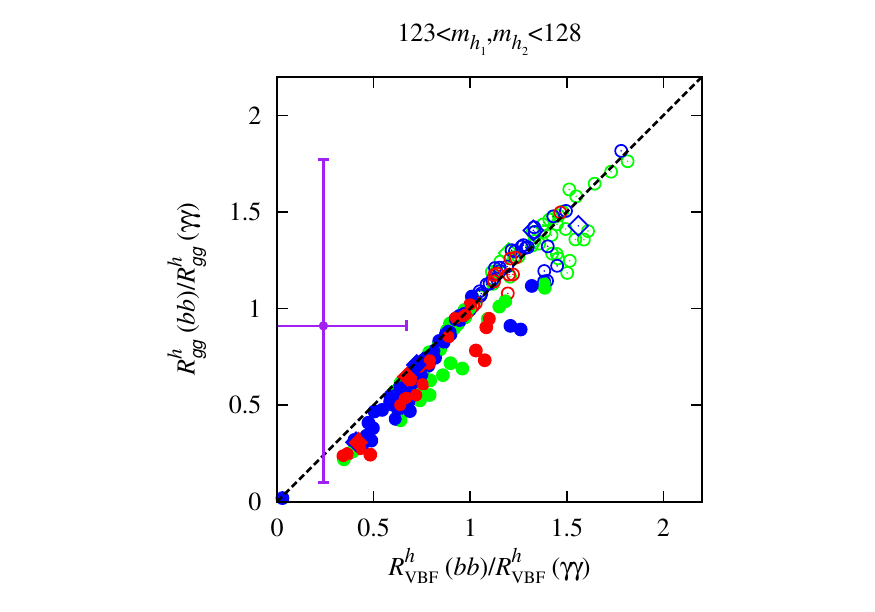}
\caption{Comparisons of pairs of event rate ratios that should be equal if only a single Higgs boson is present. The color code is green for points with $2\gev<\mhii-\mhi\leq 3\gev$, blue for $1\gev <\mhii-\mhi\leq 2\gev$, and red for $\mhii-\mhi\leq 1\gev$. Large diamond points have $\omghsq$ in the WMAP window of $[0.094,0.136]$, while circular points have $\omghsq<0.094$. Solid points are those with $\ggpp>1$ and open symbols have $\ggpp\leq 1$.  Current experimental values for the ratios  from CMS data along with their $1\sigma$ error bars are also shown. \label{plots1}}
\end{figure}

\begin{figure}[h]\centering
\includegraphics[width=0.55\textwidth]{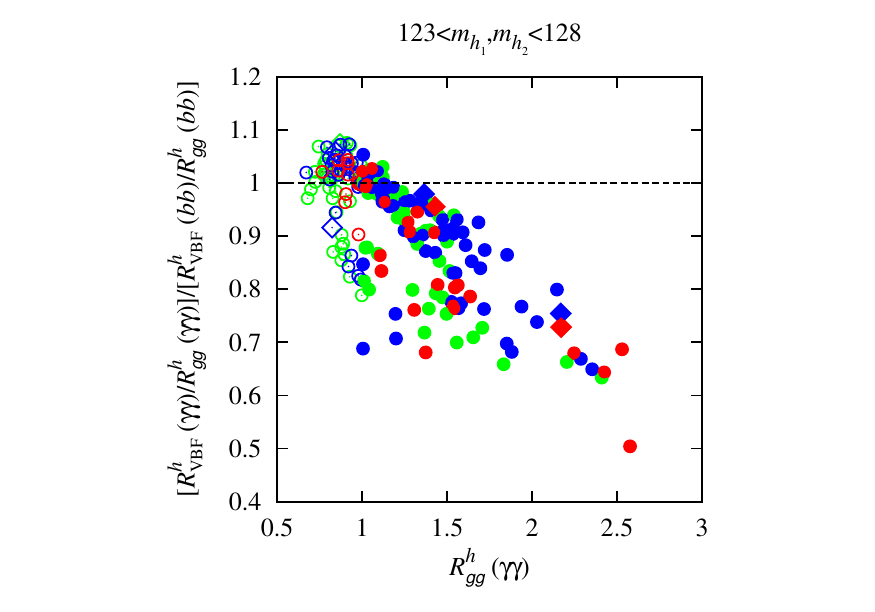}\hspace{-30mm}
\includegraphics[width=0.55\textwidth]{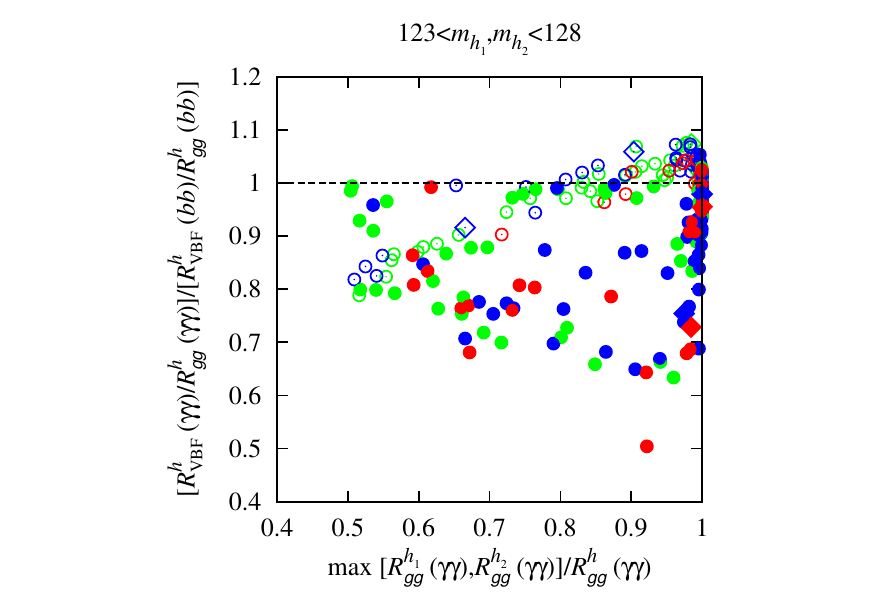}
\includegraphics[width=0.55\textwidth]{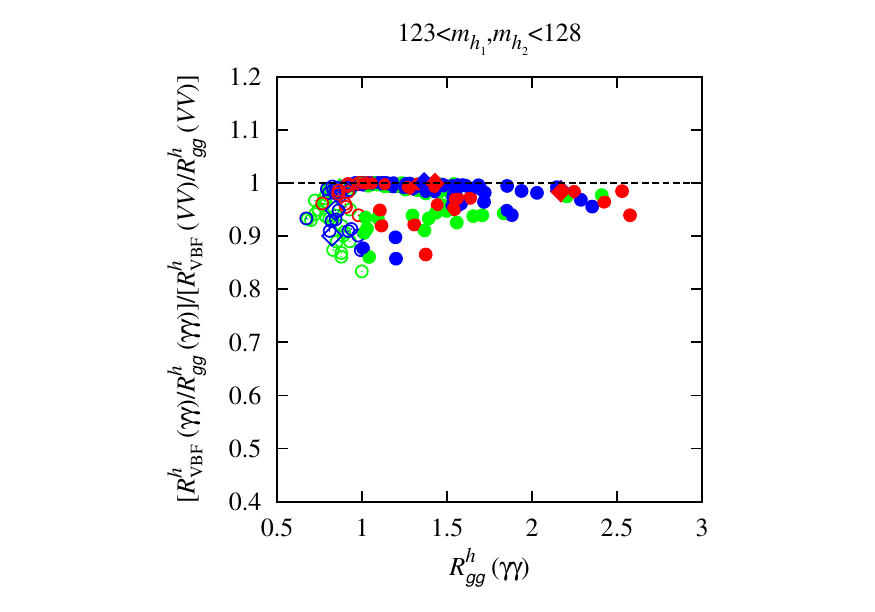}\hspace{-30mm}
\includegraphics[width=0.55\textwidth]{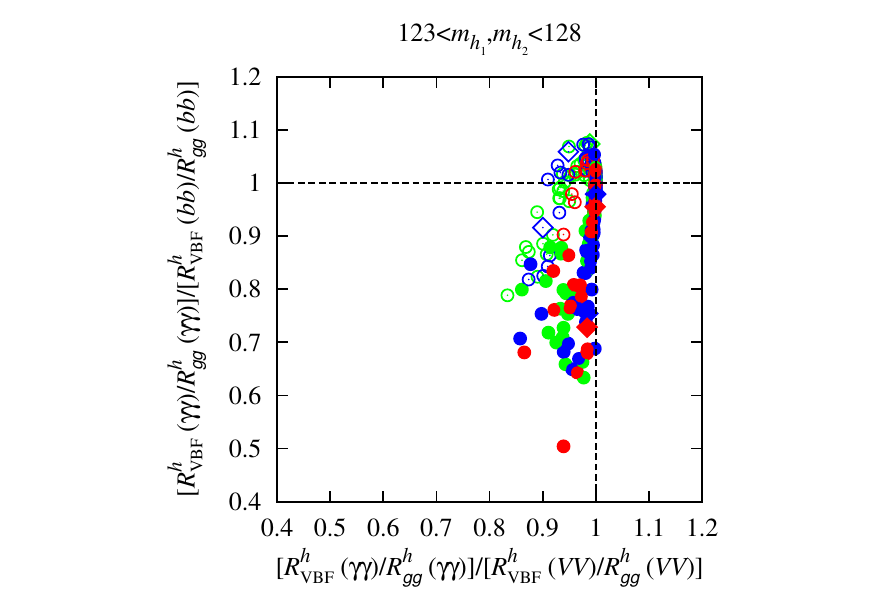}
\caption{Double ratios I) and II) of Eq.~(\ref{doublerr}) as functions of $R^h_{gg}(\gam\gam)$ (on the left).  On the right we show (top) double ratio I) vs. ${\rm max}\left[R^{\hi}_{gg}(\gam\gam),R^{\hii}_{gg}(\gam\gam)\right]/R^h_{gg}(\gam\gam)$ and (bottom) double ratio I)  vs.\ double ratio II) 
for the points displayed in Fig.~\ref{plots1}.  Colors and symbols are the same as in Fig.~\ref{plots1}. \label{plots2}}
\end{figure}

In Fig.~\ref{plots1}, we plot the numerator versus the denominator of the double ratios I) and II), III) being very like I) due to the correlation between the $\ggpp$ and $\ggww$ values discussed in  \cite{Gunion:2012gc}. We observe that any one of these double ratios will often, but not always, deviate from unity (the diagonal dashed line in the figure). 
The probability of such deviation increases dramatically if we require (as apparently preferred by LHC data)
$\ggpp>1$, see the solid (vs.\ open) symbols of Fig.~\ref{plots1}. This is further elucidated in Fig.~\ref{plots2} where we display the double ratios I) and II) as functions of 
$R^h_{gg}(\gam\gam)$ (left plots). For the NMSSM, it seems that the double ratio I)  provides the greatest discrimination between degenerate vs.\ non-degenerate scenarios with values very substantially different from unity (the dashed line) for the majority of the degenerate NMSSM scenarios explored in \cite{Gunion:2012gc} that have enhanced $\gam\gam$ rates. Note in particular that I), being sensitive to the $b\anti b$ final state,  singles out degenerate Higgs scenarios even when one or the other of $\hi$ or $\hii$ dominates the $gg\to \gam\gam$ rate, see the top right plot of Fig.~\ref{plots2}. In comparison, double ratio II) is most useful for scenarios with $R^h_{gg}(\gam\gam)\sim 1$, as illustrated by the bottom left plot of Fig.~\ref{plots2}. Thus, as illustrated by the bottom right plot of Fig.~\ref{plots2}, the greatest discriminating power is clearly obtained by measuring both double ratios. In fact, a close examination reveals that there are no points for which {\it both} double ratios are exactly 1!\footnote{We have quantitatively evaluaed the diagnostic power of the double ratios of Eq.~(\ref{doublerr}) in the Higgs-radion mixing model and again find that they deviate by substantial, often large amounts relative to unity.  In addition, there are other double ratios (see footnote \ref{f1}) that also have similar discriminating power as well as the ability to detect the custodial symmetry violation implicit in the Higgs-radion mixing model. Details will be presented in \cite{future}.}
 Of course, experimental errors may lead to a region containing a certain number of points in which both double ratios are merely consistent with 1 within the errors.

What does current LHC data say about these various double ratios?  
The central values and $1\sigma$ error bars\footnote{For the ratio $R_i/R_j$, we use 
$\sigma^{\text{upp,low}}={R_i \over R_j}\sqrt{({\sigma_i^{\text{upp,low}} / R_i})^2+({\sigma_j^{\text{upp,low}} / R_j})^2}$
to calculate its combined asymmetric $1\sigma$ error bar, where $\sigma_i^{\text{upp/low}}$ is the upper/lower $1\sigma$  
error for the individual $R_i$.} for the numerator and denominator  of double ratios I) and II) obtained from CMS data~\cite{CMS-PAS-HIG-12-020} are also shown in Fig.~\ref{plots1}.
Obviously, current statistics are inadequate to discriminate whether or not the double ratios deviate from unity. 
For a  $\rts =14\tev$ run with $L=100\fbi$ ($300\fbi$) of accumulated luminosity the SM Higgs cross sections at the relevant energies imply that the number of Higgs events will be about a factor of 25 (77) larger than the number produced for $L\sim 5\fbi$ at $\rts =7\tev$ plus $L\sim 6 \fbi$ at $\rts=8\tev$ (as used in computing the error bars shown in Fig.~\ref{plots1}). Using statistical scaling only that means the error bars plotted in  Fig.~\ref{plots1} should be reduced by roughly a factor of 5 (9),  levels that could indeed reveal a deviation from unity, or at least remove some model points if no deviation within that error is seen. Of course improvements in the experimental analyses may further increase the sensitivity.  
We thus conclude that our diagnostic tools will ultimately prove viable and perhaps crucial for determining if the $\sim 125\gev$ Higgs signal is really only due to a single Higgs-like resonance or if two resonances are contributing, the latter having significant probability in model contexts if enhanced $\gam\gam$ rates are indeed confirmed at higher statistics.


{\bf To summarize}, we have emphasized the possibility that a $\gam\gam$ Higgs-like signal that is significantly enhanced relative to the SM could arise as a result of there being two fairly degenerate Higgs bosons 
near 125~GeV. This situation arises in several model contexts in which the degeneracy can be  such that separate mass peaks could not be observed in even the high-resolution $\gam\gam$ and $ZZ\to 4\ell$ final states.
We have shown that deviations from unity of certain double ratios of event rates have strong potential for revealing the presence of two (or more) nearly degenerate Higgs bosons within the $125\gev$ LHC signal. 
Such deviations arise when both the quasi-degenerate Higgses contribute significantly to at least one production/decay channel. 
We have employed the NMSSM as a prototype model to illustrate the discriminating power of these double ratios in the case of a doublets-plus-singlets Higgs model. We have also noted that the diagnostic power of the double ratios discussed in this letter is at least as great in the brane model with Higgs-radion mixing and, in addition, there are more double ratios that can be defined as a result of the sometimes substantial violation of custodial symmetry in the latter type of model. 
Of course, substantial statistics will be required to reveal the deviations from unity that would signal a degenerate scenario.

\section*{Acknowledgements} 

This work has been supported in part by US DOE grant DE-FG03-91ER40674 and by  
IN2P3 under contract PICS FR--USA No.~5872. JFG and SK acknowledge the hospitality and inspiring working atmosphere  
of the Aspen Center for Physics which is supported by  National Science Foundation Grant No.\ PHY-1066293.


\end{document}